%% file: plenary_flavorlattice_12.tex
\documentclass[12pt]{article}

\include{econfmacros}

\usepackage[english]{babel}
\usepackage{a4,epsf,epsfig,graphicx,latexsym}

\makeatletter
\usepackage{verbatim}

\newcommand{\bea}{\begin{eqnarray}}
\newcommand{\eea}{\end{eqnarray}}
\newcommand{\nn}{\nonumber}
\newcommand{\tev}{{\rm TeV}}
\newcommand{\gev}{{\rm GeV}}
\newcommand{\mev}{{\rm MeV}}

\def\simge{\mathrel{\rlap{\raise 0.511ex \hbox{$>$}}{\lower 0.511ex
 \hbox{$\sim$}}}}
\def\simle{\mathrel{\rlap{\raise 0.511ex \hbox{$<$}}{\lower 0.511ex
 \hbox{$\sim$}}}}
\def\slash#1{\setbox0=\hbox{$#1$}\dimen0=\wd0 \setbox1=\hbox{/} \dimen1=\wd1
 \ifdim\dimen0>\dimen1 \rlap{\hbox to \dimen0{\hfil/\hfil}} #1
 \else \rlap{\hbox to \dimen1{\hfil$#1$\hfil}} / \fi}

\newcommand\pubdate{\today}

\def\roma{Dipartimento di Fisica, Universit\`a Roma Tre and INFN, Sezione di Roma Tre,\\
        Via della Vasca Navale 84, I-00146 Roma, Italy}

\def\Title#1{\begin{center} {\Large #1 } \end{center}}
\def\Author#1{\begin{center}{ \sc #1} \end{center}}
\def\Address#1{\begin{center}{ \it #1} \end{center}}

\newcommand\pubblock{\rightline{\begin{tabular}{l} 
\pubdate  \end{tabular}}}
\newenvironment{Abstract}{\begin{quotation}  }{\end{quotation}}
\newenvironment{Presented}{\begin{quotation} \begin{center} 
             PRESENTED AT\end{center}\bigskip 
      \begin{center}\begin{large}}{\end{large}\end{center} \end{quotation}}
\def\Acknowledgements{\bigskip  \bigskip \begin{center} \begin{large}
             \bf ACKNOWLEDGEMENTS \end{large}\end{center}}
\input econfmacros.tex

\begin{document}
\begin{titlepage}
\pubblock

\vfill
\Title{Flavor Lattice QCD in the Precision Era}
\vfill
\Author{ Cecilia Tarantino}
\Address{\roma}
\vfill
\begin{Abstract}
I discuss the important role that Lattice QCD plays in testing the Flavor sector of the Standard Model (SM) and in indirect searches of New Physics. I review in particular the Unitarity Triangle Analysis performed by the UTfit collaboration within and beyond the SM, presenting the recent accurate lattice results that enter the analyses. I conclude with a tentative outlook to the further progresses that we can expect in the next years from Flavor Lattice QCD.
\end{Abstract}
\vfill
\begin{Presented}
the $5^{th}$ International Workshop on Charm Physics,\\ Charm2012, May 14-17 2012, Honolulu (Hawaii);\\
the $30^{th}$ International Symposium on Lattice Field Theory,\\ Lattice2012, June 24-29 2012, Cairns (Australia);\\
the $36^{th}$ International Conference on High Energy Physics,\\ ICHEP2012, July 4-11 2012, Melbourne (Australia).

\end{Presented}
\vfill
\end{titlepage}
\def\thefootnote{\fnsymbol{footnote}}
\setcounter{footnote}{0}
%
%
%
%
%
%
%



\section{Introduction}
\label{sec:intro}
This year, 2012, marks the beginning of a new era in Physics. A new boson has been observed by Atlas and CMS, at a mass of approximately 125 GeV~\cite{atlas,cms}, that is in the range where the Higgs boson, which is the last missing Standard Model particle, is expected.
Studies on the nature of this particle will be performed aiming at understanding if the Standard Model (SM) is all what we can see in present experiments or if New Physics (NP) effects may be revealed.

In order to search for NP and understand its nature there is a research activity that is complementary to the direct production of NP particles, that is Flavor Physics. Studies of Flavor Physics look at theoretically clean and SM suppressed processes where NP effects may be competitive to the SM contributions and thus visible.
Typically, the dominant uncertainty in the theoretical predictions of flavor observable is brought by the hadronic parameters which enclose the (non-perturbative) long-distance QCD contributions. It is therefore crucial to have accurate computations of the hadronic parameters. A leading role is played by Lattice QCD as it is a non-perturbative approach based on first principles. It consists in simulating QCD itself, without any additional parameter, on a discrete space-time and in a finite volume.

Lattice QCD has recently entered the precision era thanks to the increased computational power and the algorithm and action improvements achieved in the last decade. The former has led to the so-called unquenched calculations, where the contribution of loops of dynamical quarks is taken into account. In the last decade essentially all lattice calculations have been performed with either two (up and down) or three (up, down and strange) dynamical quarks. Some very recent calculations also include the contribution of the dynamical charm quark~\cite{Baron:2010bv,Baron:2010th,Bazavov:2010ru}. Thanks to the algorithm and action improvements, simulations at light quark masses in the Chiral Perturbation Theory (ChPT) regime have become feasible and, very recently, first simulations at the physical point have been performed~\cite{Durr:2010aw,Aoki:2009ix}.

A clear indication of the level of accuracy achieved at present in Flavor Lattice QCD calculations is given by the color code introduced by the Flavor Lattice Averaging Group (FLAG)~\cite{flag}. The task of FLAG is to review lattice results of interest for Flavor Physics and to provide lattice averages, which include lattice results where all systematic uncertainties are satisfactorily under control. More in detail, a color tag is assigned to the lattice results w.r.t. each systematic uncertainty. Green, orange and red colors respectively correspond to the cases of a systematic uncertainty that is completely, sufficiently or not enough under control. Lattice results have to have no red tags to be included in the average.
The first FLAG review~\cite{Colangelo:2010et} provided averages for pion and kaon Physics. A second review updating the previous one and including also Heavy Flavor hadronic parameters is in progress~\cite{flagnew}.
A green tag is assigned for the continuum extrapolation if the analysis has been performed with at least three lattice spacings with at least two values below $0.1$ fm. The condition for a green tag for the chiral extrapolation is that the simulated pion masses are lighter than 250 MeV. The renormalization, where needed, has to be non-perturbative for a green tag. Finite volume effects are considered to be completely under control if the product $M_{\pi, min} \cdot L > 4$ or at least three volumes are simulated.
The chosen criteria, which reflect the state of the art of present lattice results, provide evidence of the high level of accuracy achieved in lattice calculations.

In the following, in order to show the important role of Lattice QCD in Flavor Physics, I will discuss an emblematic analysis that relies on several lattice results: the determination of the parameters of the Cabibbo-Kobayashi-Maskawa matrix and in particular the Unitarity Triangle Analysis performed by the UTfit collaboration.

\section{The Cabibbo-Kobayashi-Maskawa matrix}
One of the main tasks of Flavor Physics is an accurate determination of the parameters of the Cabibbo-Kobayashi-Maskawa (CKM) matrix. It represents a crucial test of the SM and, moreover, improving the accuracy on the CKM parameters is at the basis of NP analyses, where small NP effects are looked for.
The CKM matrix, $V_{CKM}$, is the mixing matrix that relates (primed) weak eigenstates to (unprimed) mass eigenstates for down-type quarks, as follows
\bea
\left(\begin{array}{c} d^\prime\\s^\prime\\b^\prime \end{array}\right) = V_{CKM}\, \left(\begin{array}{c} d\\s\\b \end{array}\right)= \left( \begin{array}{ccc} V_{ud} & V_{us} & V_{ub}\\
V_{cd} & V_{cs} & V_{cb}\\
V_{td} & V_{ts} & V_{tb}
 \end{array} \right)\,\, \left(\begin{array}{c} d\\s\\b \end{array}\right)\,.
\label{eq:ckm}
\eea
In the mass eigenstate basis, therefore, the CKM matrix elements appear in weak charged currents.
Being a 3x3 unitarity matrix, $V_{CKM}$ depends on four independent physical parameters: three mixing angles and one phase. In the convenient Wolfenstein parameterization, the CKM matrix is expressed in terms of the four parameters A, $\lambda$, $\rho$ and $\eta$, as an expansion in the small parameter $\lambda$, which is the sine of the Cabibbo angle ($\lambda=\sin \theta_c \approx 0.2$).
Up to $\mathcal{O}(\lambda^5)$, as required by the present level of theoretical and experimental accuracy, the Wolfenstein parameterization of the CKM matrix reads
\bea
\footnotesize{
V_{CKM} = \left( \begin{array}{ccc} 1-\frac{1}{2}\lambda^2-\frac{1}{8}\lambda^4 & \lambda & A\,\lambda^3\,(\rho-i\, \eta)\\
-\lambda+\frac{1}{2}A^2\,\lambda^5[1-2(\rho+i\, \eta)] & 1-\frac{1}{2}\lambda^2-\frac{1}{8}\lambda^4 (1+4\,A^2) & A\,\lambda^2\\
A\,\lambda^3\,[1-(\rho+i\, \eta)(1-\frac{1}{2}\lambda^2)]  & -A\,\lambda^2+\frac{1}{2} A (1-2\,\rho)\lambda^4 - i\,\eta\, A\, \lambda^4 &  1-\frac{1}{2}A^2\,\lambda^4
 \end{array} \right)\,.
 }
\label{eq:wolfenstein}
\eea

In the determination of the four CKM parameters Lattice QCD plays a crucial role.
The parameter $\lambda$, besides being the CKM expansion parameter, it is particularly interesting as it enters the most stringent unitarity condition on the CKM matrix.
This is, among the nine unitarity conditions $V_{CKM}^\dagger V_{CKM}=1$, the first-row relation which reads $|V_{ud}|^2+ |V_{us}|^2+ |V_{ub}|^2=1$.
In this relation the contribution of $|V_{ub}|^2$ can be neglected as it is very small, at the level of and even  slightly smaller than present uncertainties on $|V_{ud}|^2$ and $|V_{us}|^2$.
The present uncertainty on the first row unitarity condition is almost equally distributed between $|V_{ud}|^2$ ($4\cdot 10^{-4}$) and $|V_{us}|^2$ ($5\cdot 10^{-4}$).
The parameter $|V_{ud}|$ is very precisely determined, at the $0.02\%$ level, from nuclear beta decays.
It can be alternatively determined, at a similar level of accuracy, from the leptonic pion decay, relying on the lattice computation of the pion decay constant.
The parameter $|V_{us}|$ relies on the Lattice results for the kaon decay constant $f_K$ or for the vector form factor $f_+(0)$. The former results allow to determine $|V_{us}|$ from the experimental measurement of the so-called $Kl2$ leptonic decay $K \to \mu \nu$~\cite{Marciano:2004uf}, while the latter results are required to extract $|V_{us}|$ from the experimental measurement of the so-called $Kl3$ semileptonic decay $K \to \pi l \nu$~\cite{Leutwyler:1984je,Becirevic:2004ya}.
For a recent review of the $f_K$ and $f_+(0)$ lattice results I refer to the FLAG review~\cite{flag} and to Gilberto Colangelo's plenary talk at the Lattice2012 conference~\cite{colangelopos}.
Here I only quote the FLAG averages for $|V_{us}|$, which combine both the $Kl2$ and $Kl3$ determinations and are separately given for the $N_f=2$ and $N_f=2+1$ lattice input: $|V_{us}|=0.2254 \pm 0.0009$ from $N_f=2+1$ and $|V_{us}|=0.2251 \pm 0.0018$ from $N_f=2$.

As $|V_{us}|$ is known at present with the impressing precision of $0.5\%$, small effects of the same sub-percent size, like isospin breaking (IB) effects, have now to be included in the determination.
So far lattice calculations have been typically performed in the limit of exact isospin symmetry, that is with degenerate up and down quark masses ($m_u=m_d$) and neglecting electromagnetic effects ($Q_u=Q_d=0$). The parametric size of the IB effects is of approximately 1\% as they are of $\mathcal{O}(\alpha_{e.m.})$ or $\mathcal{O}((m_d-m_u)/\Lambda_{QCD})$ depending on the electromagnetic ($Q_u\neq Q_d$) or strong interaction ($m_u\neq m_d$) origin.
Last year, the strong IB corrections to $f_K/f_\pi$ and to $f_+(0)$ have been calculated on the Lattice for the first time~\cite{deDivitiis:2011eh}\footnote{The strong IB effects were previously taken into account in the analysis of~\cite{Aubin:2004fs} by fitting isospin symmetric lattice data through ChPT formulas.}. The study of ref.~\cite{deDivitiis:2011eh} is not performed removing directly the degeneracy $m_u=m_d$, it is instead based on the idea of expanding the functional integral in the small parameter $(m_d-m_u)/\Lambda_{QCD}$ up to first order, with the advantage of computing the (not small) slope in $(m_d-m_u)/\Lambda_{QCD}$. By comparing one lattice result, for instance for the kaon mass splitting, to the corresponding experimental value, the quark mass splitting $(m_d-m_u)$ turns out to be determined. The results obtained in~\cite{deDivitiis:2011eh} by adopting this {\it expansion} method read
\bea
\left[m_d-m_u\right]^{QCD}(\overline{MS},2 GeV) &=& 2.35(8)(24)\ \mbox{MeV} 
\quad \times \quad 
\frac{\left[M_{K^0}^2-M_{K^+}^2\right]^{QCD}}{6.05\times 10^3\ \mbox{MeV}^2} \, ,
\nonumber \\
\nonumber \\
\left[\frac{F_{K^+}/F_{\pi^+}}{F_{K}/F_{\pi}}-1\right]^{QCD} &=& -0.0039(3)(2)
\quad \times \quad 
\frac{\left[M_{K^0}^2-M_{K^+}^2\right]^{QCD}}{6.05\times 10^3\ \mbox{MeV}^2} \, ,
\label{eq:mainresults}
\eea
where $[M_{K^0}^2-M_{K^+}^2]^{QCD}$ represents the kaon mass splitting due to strong IB effects.
In these results the lattice error (the one in the first parenthesis) has been obtained with a rather modest statistics,  $\sim 150$ gauge field configurations, and is expected to be reduced in the next future. Most of the systematic error (the one in the second parenthesis) comes from the ambiguity in the definition of the electromagnetic corrections in the experimental input $\left[M_{K^0}^2-M_{K^+}^2\right]^{QCD}$, which is also expected to be reduced thanks to future lattice computations of the complementary electromagnetic effect~\cite{izubuchi,ibem}.

\section{The UTA within the Standard Model}
The CKM parameters $\rho$ and $\eta$ are conveniently determined through the so-called Unitarity Triangle Analysis (UTA)~\cite{Ciuchini:2000de}-\cite{utfit}, which consists in constraining sides and angles of the triangle defined in the $(\overline{\rho},\overline{\eta})$-plane ($\overline{\rho} \equiv \rho (1-\lambda^2/2)$ and $\overline{\eta} \equiv \eta (1-\lambda^2/2)$) by the unitarity condition $V_{ub}^* V_{ud}+V_{cb}^* V_{cd}+V_{tb}^* V_{td}=0$ which involves the first and third rows of the CKM matrix.
This triangle has the advantage of having sides of similar size and thus of being sensitive to the CP-violating parameter $\overline{\eta}$.

Within the UTA several constraints are included, which are provided by the comparison between experimental measurement and theoretical prediction for flavor observables that depend on $\overline{\rho}$ and $\overline{\eta}$.
The list of the constraints and their present level of accuracy is given in table~\ref{tab:constraints}. 
\begin{table}[h!]
\begin{center}
\begin{tabular}{||c|c||}
\hline
Observable & Accuracy \\
\hline
$\varepsilon_K$ & $0.5$\%\\
$\Delta m_d$ & $1$\%\\
$\Delta m_d/\Delta m_s$ & $1$\%\\
$|V_{ub}/V_{cb}|$ & $15$\%\\
$Br(B \to \tau \nu)$ & $20$\%\\
$\sin 2 \beta$ & $3$\%\\
$\cos 2 \beta$ & $15$\%\\
$\alpha$ & $7$\%\\
$\gamma$ & $14$\%\\
$2\beta+\gamma$ & $50$\%\\
\hline
\end{tabular}
\end{center}
\caption{Approximate level of accuracy on the UTA constraints.}
\label{tab:constraints}
\end{table}
\begin{figure}[th!]
\begin{center}
\vspace{-6.0cm}
\includegraphics[scale=0.45]{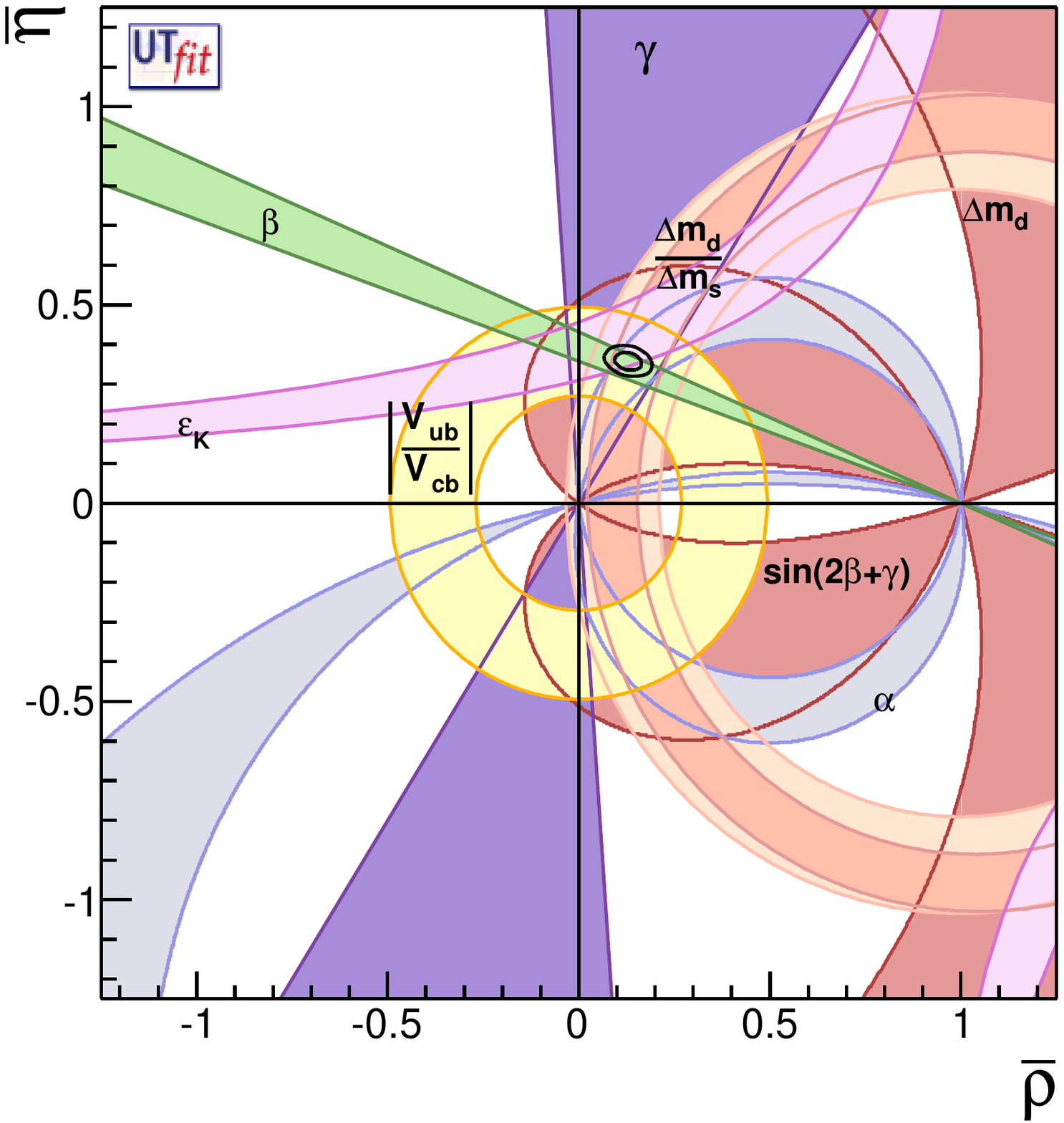} \\
\end{center}
\vspace{-0.5cm}
\caption{\sl Results of the UTA within the SM. The contours display the selected 68\% and 95\% probability regions in the $(\overline{\rho},\overline{\eta})$-plane. The 95\% probability regions selected by the single constraints are also shown.
}
\label{fig:main}
\end{figure}
For some constraints the experimental accuracy is at the level of few percent or even better, so that a significant comparison to the theoretical prediction calls for a similarly good control of the hadronic uncertainties.
The hadronic parameters required in the UTA are the bag-parameter $B_K$ entering the theoretical prediction of $\varepsilon_K$, the semileptonic form factors $f_+$ and $F$ required for the extraction of $|V_{ub}|$ and $|V_{cb}|$ and the combinations of $B_{(s)}$-meson decay constants and bag parameters $f_{Bs}$, $f_{Bs}/f_B$, $B_{Bs}$ and $B_{Bs}/B_B$, which enter the theoretical predictions of the B-physics observables $\Delta m_d$, $\Delta m_d/\Delta m_s$ and $Br(B\to \tau \nu)$. 
\begin{table}[t!]
\begin{center}
\begin{tabular}{||c|c|c|c||}
\hline
Observable & Input value & SM prediction & Pull \\
\hline
$\varepsilon_K\cdot 10^3$ & $2.23\pm0.01$ & $1.96\pm0.20$ & $1.4$ \\
$\Delta m_s [ps^{-1}]$ & $17.69\pm 0.08$ & $18.0\pm 1.3$ & $<1$ \\
$|V_{cb}|\cdot 10^3$ & $41.0\pm1.0$ & $42.3\pm0.9$ & $<1$ \\
$|V_{ub}|\cdot 10^3$ & $3.82\pm0.56$ & $3.62\pm 0.14$ & $<1$\\
$Br(B \to \tau \nu)\cdot 10^4$ & $1.67\pm0.30$ & $0.82\pm 0.08$ & $2.7$ \\
$\sin 2 \beta$ & $0.68\pm0.02$ & $0.81\pm0.05$ & $2.4$ \\
$\alpha$ & $91^\circ\pm6^\circ$ & $88^\circ\pm4^\circ$ & $<1$ \\
$\gamma$ & $76^\circ\pm 11^\circ$ & $68^\circ\pm 3^\circ$ & $<1$ \\
\hline
\end{tabular}
\end{center}
\caption{Comparison between input value and SM prediction for the UTA constraints. The pull is also shown.}
\label{tab:pull}
\end{table}

\begin{figure}[t!]
\begin{center}
\vspace{-1.7cm}
\includegraphics[scale=0.35]{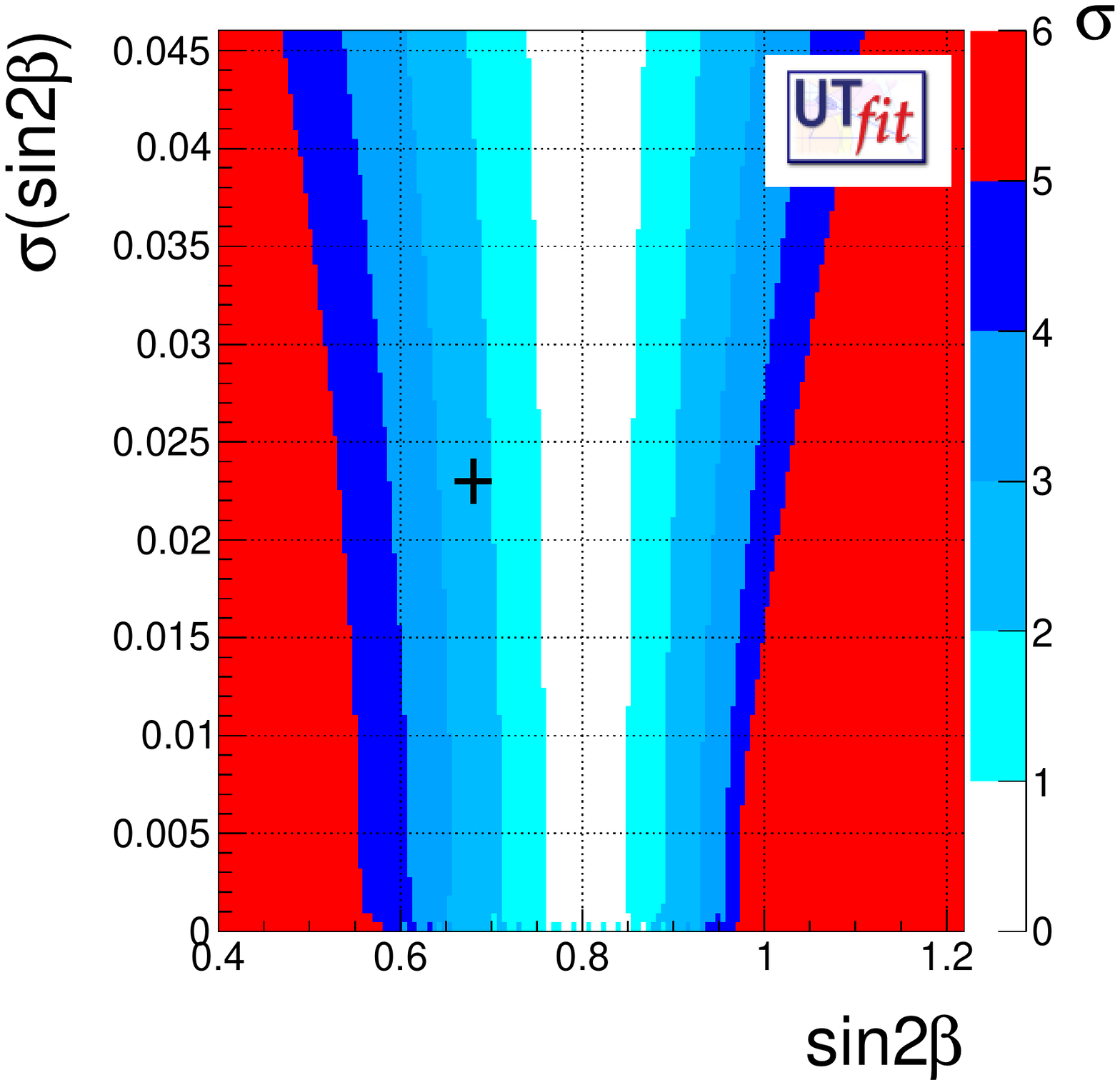} \\
\end{center}
\vspace{-0.7cm}
\caption{\sl Compatibility plot for $\sin 2 \beta$. The input value is $2.4 \sigma$ smaller than the UTA prediction.}
\label{fig:sin2b}
\end{figure}

The main results of the UTA~\cite{utfit}, performed by the UTfit collaboration assuming the validity of the SM, are summarized in fig.~\ref{fig:main}, where the curves representing the UTA constraints intersect in a single allowed region for $(\overline{\rho},\overline{\eta})$, proofing that the CKM parameters are consistently overconstrained.
In other words, the UTA has established that the CKM matrix is the dominant source of flavor mixing and CP violation and the parameters $\overline{\rho}$ and $\overline{\eta}$ turn out to have the values $\overline{\rho}=0.139 \pm 0.021$ and $\overline{\eta}=0.352\pm0.016$.
\begin{figure}[t!]
\begin{center}
\vspace{-2.0cm}
\includegraphics[scale=0.35]{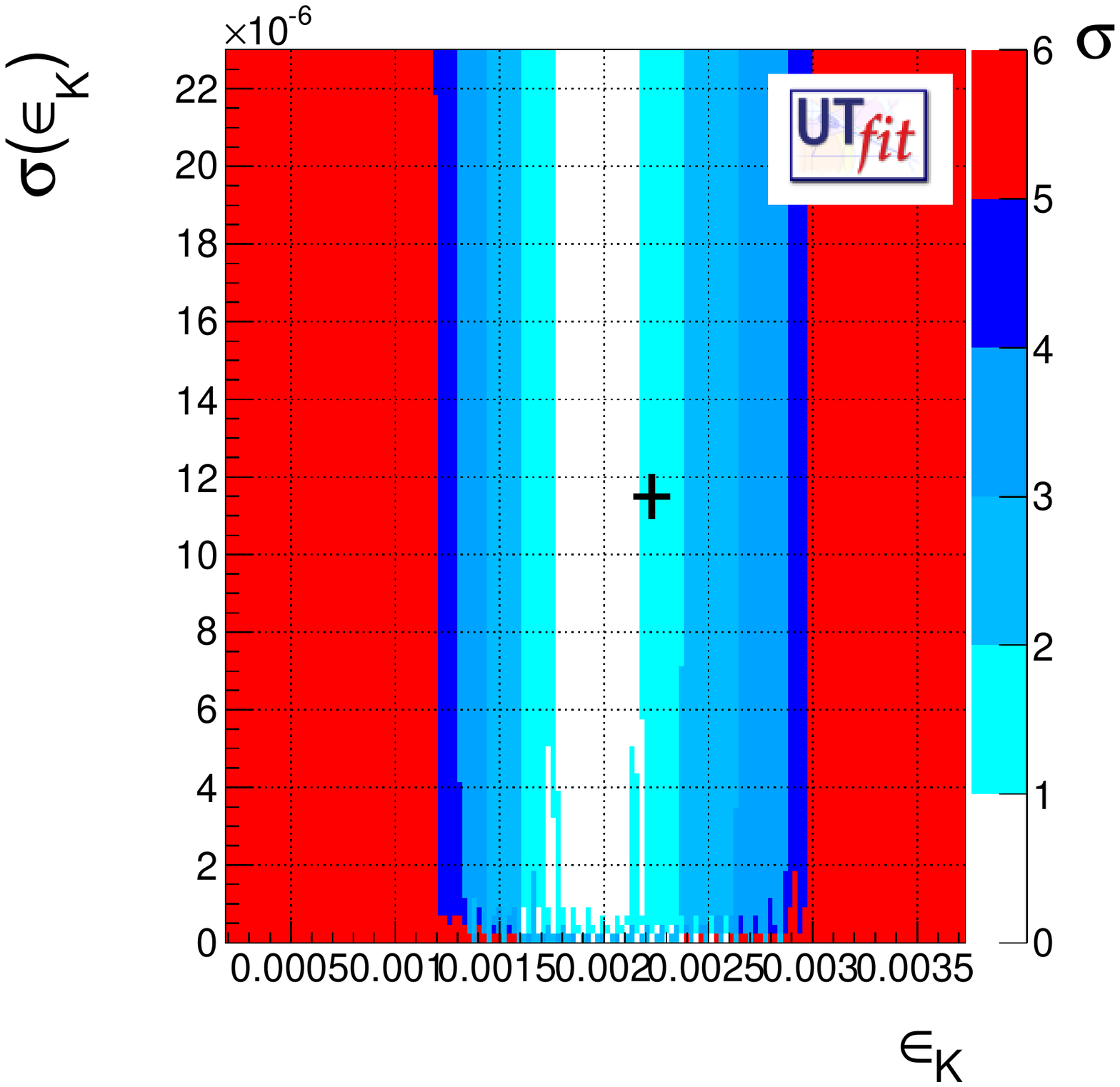} \\
\end{center}
\vspace{-0.7cm}
\caption{\sl Compatibility plot for $\varepsilon_K$. The cross, which represents the input value, is $1.4 \sigma$ larger than the UTA prediction.}
\label{fig:epsK}
\end{figure}
\begin{figure}[h!]
\begin{center}
\vspace{-1.7cm}
\includegraphics[scale=0.35]{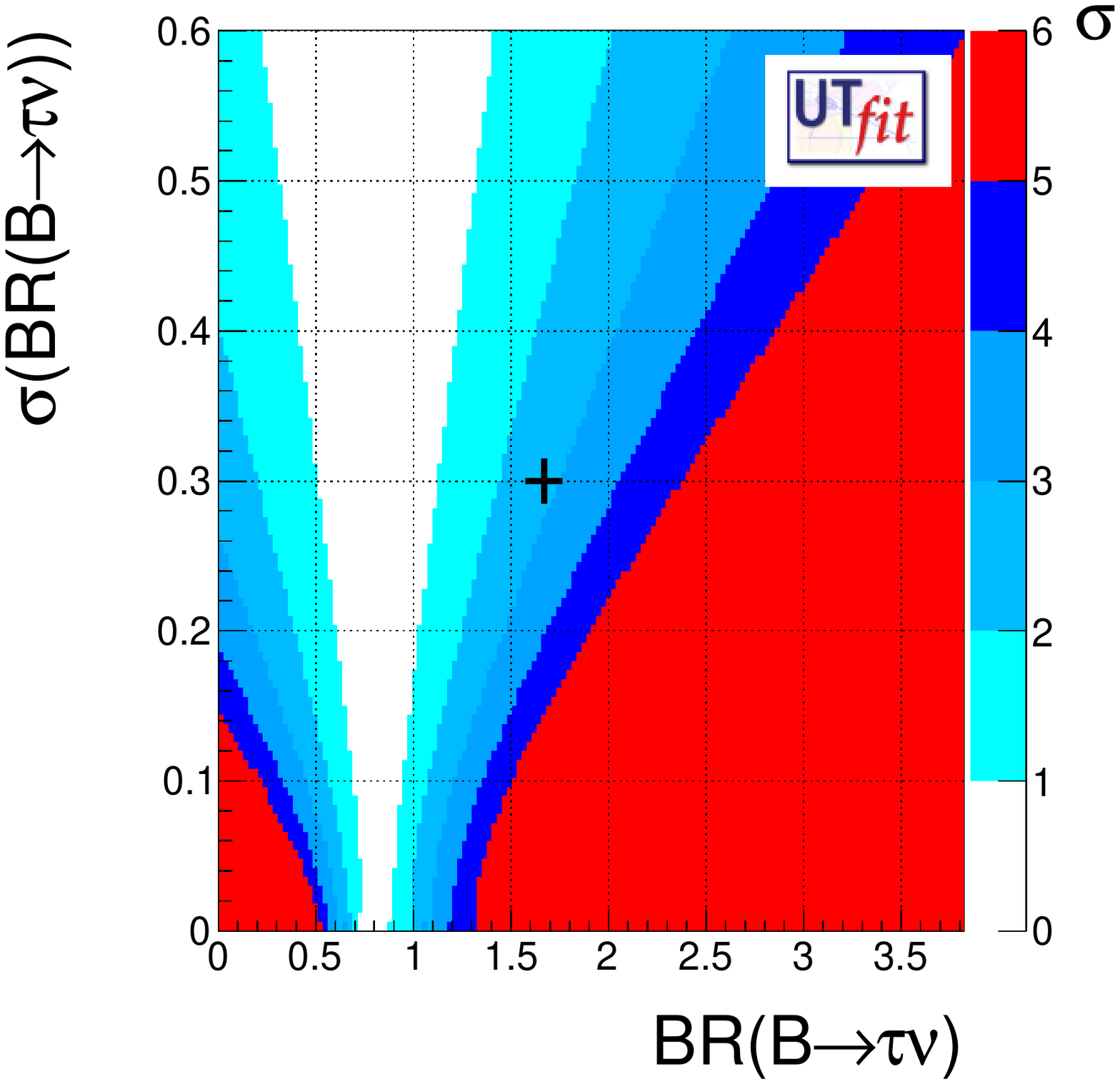} \\
\end{center}
\vspace{-0.7cm}
\caption{\sl Compatibility plot for $Br(B \to \tau \nu)$. The input value is $2.7 \sigma$ larger than the UTA prediction.}
\label{fig:Btaunu}
\end{figure}
In table~\ref{tab:pull} this comparison is shown for all the UTA constraints and the pull (i.e. the difference between the input value and the SM UTA prediction divided by the uncertainty) is provided as well.
For most of the constraints the pull is smaller than one, showing that there is a very good compatibility between the input value and the UTA prediction. For the three observables $\varepsilon_K$, $Br(B \to \tau \nu)$ and $\sin 2 \beta$, instead, there is some tension as shown by the pull that is larger than unity and by the compatibility plots in figs.~\ref{fig:epsK}-\ref{fig:sin2b}.

The theoretical prediction for the CP-violating parameter $\varepsilon_K$ depends on the bag-parameter $\hat B_K$ ($\hat B$ denotes the renormalization group invariant B-parameter) which encloses the long-distance contribution in $K^0-\bar K^0$ mixing and for which several unquenched results have recently become available.
The $N_f=2+1$ and $N_f=2$ FLAG averages read $\hat B_K^{N_f=2+1}=0.738\pm0.020$ and $\hat B_K^{N_f=2}=0.729\pm0.030$~\cite{Colangelo:2010et}. The input value adopted in the UTA is slightly larger, $\hat B_K=0.750\pm0.020$, to take into account new results~\cite{Durr:2011ap}-\cite{Kelly:2012uy} that have appeared after the FLAG review. In particular, the result in~\cite{Durr:2011ap} is characterized by a very safe chiral extrapolation since the simulated pion masses are close to the physical value, thanks to the choice of a particularly advantageous Lattice QCD action, the 2-step HEX smeared clover-improved Wilson action~\cite{Capitani:2006ni}.

The observable where the tension between experimental measurement and UTA prediction is the largest is $Br(B \to \tau \nu)$, for which the average of the BaBar and Belle experimental measurements reads $Br(B \to \tau \nu)_{exp}=(1.67 \pm  0.30) \cdot 10^{-4}$ while the UTA prediction, which assumes the SM validity, turns out to be $Br(B \to \tau \nu)_{SM}=(0.82 \pm 0.08) \cdot 10^{-4}$.
In wondering if this $2.7\sigma$ deviation can be due to NP effects the first model that comes to theorists' mind is the simplest 2-Higgs Doublet Model (2HDM), that is the 2HDM of type II, where one Higgs boson doublet ($H_u$) couples to up-type quarks and the other Higgs boson doublet ($H_d$) couples to down-type quarks. The observed deviation, in principle, could be easily explained in the 2HDM of type II where, in addition to the tree-level SM amplitude with the W-boson exchange, there is a tree-level contribution mediated by the charged Higgs. As the Higgs couples more strongly to the heavy $\tau$ lepton than to the lightest muon and electron, the 2HDM of type II would seem to provide a natural explanation of the fact that the deviation is seen in the $\tau$ channel only. 
However, in order to explain the enhancement observed in $Br(B \to \tau \nu)$, the 2HDM of type II should have a large value of $\tan \beta/m_H^+$ which is instead excluded by other constraints, in particular by the experimental measurement of $Br(b\to s \gamma)$.
At the ICHEP conference an important experimental news has been announced  by the Belle collaboration~\cite{Adachi:2012mm}. Belle has performed a new analysis, with a modified hadronic tag, finding a result for $Br(B \to \tau \nu)$ that is significantly smaller than the previous one and that is compatible with the SM prediction. Thus, the present experimental average reads $Br(B \to \tau \nu)_{exp}=(0.99 \pm  0.25) \cdot 10^{-4}$.
Further experimental results are certainly looked forward.
From the theory side one could wonder if the observed enhancement could be due  to some underestimated uncertainty instead of NP effects. The theoretical prediction for $Br(B \to \tau \nu)$ is proportional to $|V_{ub}|^2$, which represents the main source of uncertainty in the branching ratio, mainly due to the $2.6\sigma$ difference between the inclusive and the exclusive determinations of this CKM element. The experimental measurements of $Br(B \to \tau \nu)$ would prefer a large value of $|V_{ub}|$, close to the inclusive determination. However, such a large value would not solve rather would worsen the tension in $\sin 2\beta$ and, therefore, it does not seem to be the solution to the $Br(B \to \tau \nu)$ puzzle.

The interest for B decays with a $\tau$ lepton in the final state has been recently stimulated also by the new BaBar (full data) results for the two ratios $R(D^{(*)})=Br(\bar B \to D^{(*)} \tau^- \bar \nu_t)/ Br(\bar B \to D^{(*)} \ell^- \bar \nu_{\ell})$~\cite{Lees:2012xj}, which respectively exceed the SM predictions by $2.0$ and $2.7 \sigma$, corresponding to a combined discrepancy at the $3.4\sigma$ level.
In two recent papers~\cite{Becirevic:2012jf,Bailey:2012jg} a more accurate theoretical prediction of the $R(D)$ ratio has been provided. The idea of~\cite{Becirevic:2012jf} is to obtain an estimate of $R(D)$ with minimal theory input, in particular by using in input the ratio of the vector and scalar form factors. In~\cite{Bailey:2012jg}, instead, the input value for the scalar form factor is taken from unquenched Lattice QCD only~\cite{Bailey:2012rr}. Both papers~\cite{Becirevic:2012jf} and~\cite{Bailey:2012jg} slightly reduce the discrepancy of the theoretical prediction for $R(D)$ with the experimental measurement, from $2.0 \sigma$ to $1.8$ and $1.7\sigma$ respectively.
The 2HDM of type II, that as in the case of $Br(B \to \tau \nu)$ in principle could provide an explanation to the enhancements in $R(D^{(*)})$ in terms of a charged Higgs contribution, would require in this case two different values of $\tan \beta/m_H^+$ to explain the experimental results for $R(D)$ and $R(D^*)$.
More elaborated NP models, instead, could accommodate the enhancements observed in  $Br(B \to \tau \nu)$ and in $R(D^{(*)})$. Some of them are 2HDM of type III (where the $H_u$ and $H_d$ bosons couple to both up- and down-type quarks) with flavor violation in the up sector~\cite{Crivellin:2012ye} and NP models with right-right vector and right-left scalar currents, like some 2HDM, leptoquarks or composite quarks and leptons models (with non-trivial flavor structure)~\cite{Fajfer:2012jt}.

\section{B-Physics lattice inputs for the UTA}
Lattice results for B-Physics hadronic parameters play a crucial role in the UTA. Indeed, the five UTA constraints that rely on Lattice QCD results are $\varepsilon_K$, $Br(B \to \tau \nu)$, $\Delta m_d$, $\Delta m_d/ \Delta m_s$ and $|V_{ub}/V_{cb}|$, with the last four being B-Physics observables.

The computation of B-Physics hadronic parameters on the Lattice is complicated by the presence of large discretization effects of order $(a*m_b)$ up to some power, which imply that the physical b-quark mass, being of approximately $4\gev$, cannot be directly simulated on present lattices (where $a^{-1} \leq 4 \gev$).
Several methods have been investigated and adopted so far, that are either based on an effective theory approach or  consist in simulating heavy quark masses ($m_h$) in the charm region (or slightly above) and using some suitable technique to achieve the b-quark region. These are the Tor Vergata step-scaling method~\cite{Guagnelli:2002jd}, which consists in matching several lattice simulations at different volumes and up to the physical b-quark mass on the small volume, the ETMC ratio method~\cite{Blossier:2009hg} based on suitable ratios with exactly known static limit, the HPQCD HISQ simulation~\cite{Follana:2006rc} which reduces discretization terms to $\mathcal{O}(\alpha_s a^2m_h^2)$ and $\mathcal{O}(a^4 m_h^4)$.
Approaches based on effective theory ideas are HQET and NRQCD simulations.
Within the FermiLab method~\cite{ElKhadra:1996mp} the key discretization errors are removed by tuning three parameters combining perturbation theory and experimental inputs.
Finally, this year, RBC/UKQCD has formulated the non-perturbatively tuned relativistic heavy-quark action~\cite{Aoki:2012xaa}. It is a variant of the FermiLab approach with fully non-perturbative tuning within the $B_s^{(*)}$ system.

In the following of this section I will review the state of the art of the lattice results for the B-Physics hadronic parameters entering the UTA. They are the combinations of $B_{(s)}$-meson decay constants and bag parameters $f_{Bs}$, $f_{Bs}/f_B$, $B_{Bs}$ and $B_{Bs}/B_B$, which enter the theoretical predictions of $\Delta m_d$, $\Delta m_d/\Delta m_s$ and $Br(B\to \tau \nu)$ and the semileptonic form factors $f_+$ and $F$ required for the extraction of $|V_{ub}|$ and $|V_{cb}|$. I will also quote the averages, for these hadronic parameters, that are used in the UTA by the UTfit collaboration and that represent an update w.r.t. to ref.~\cite{Lubicz:2008am}.  

\begin{figure}[t!]
\begin{center}
\begin{tabular}{cc}
\hspace*{-0.5cm}
\includegraphics[scale=0.31]{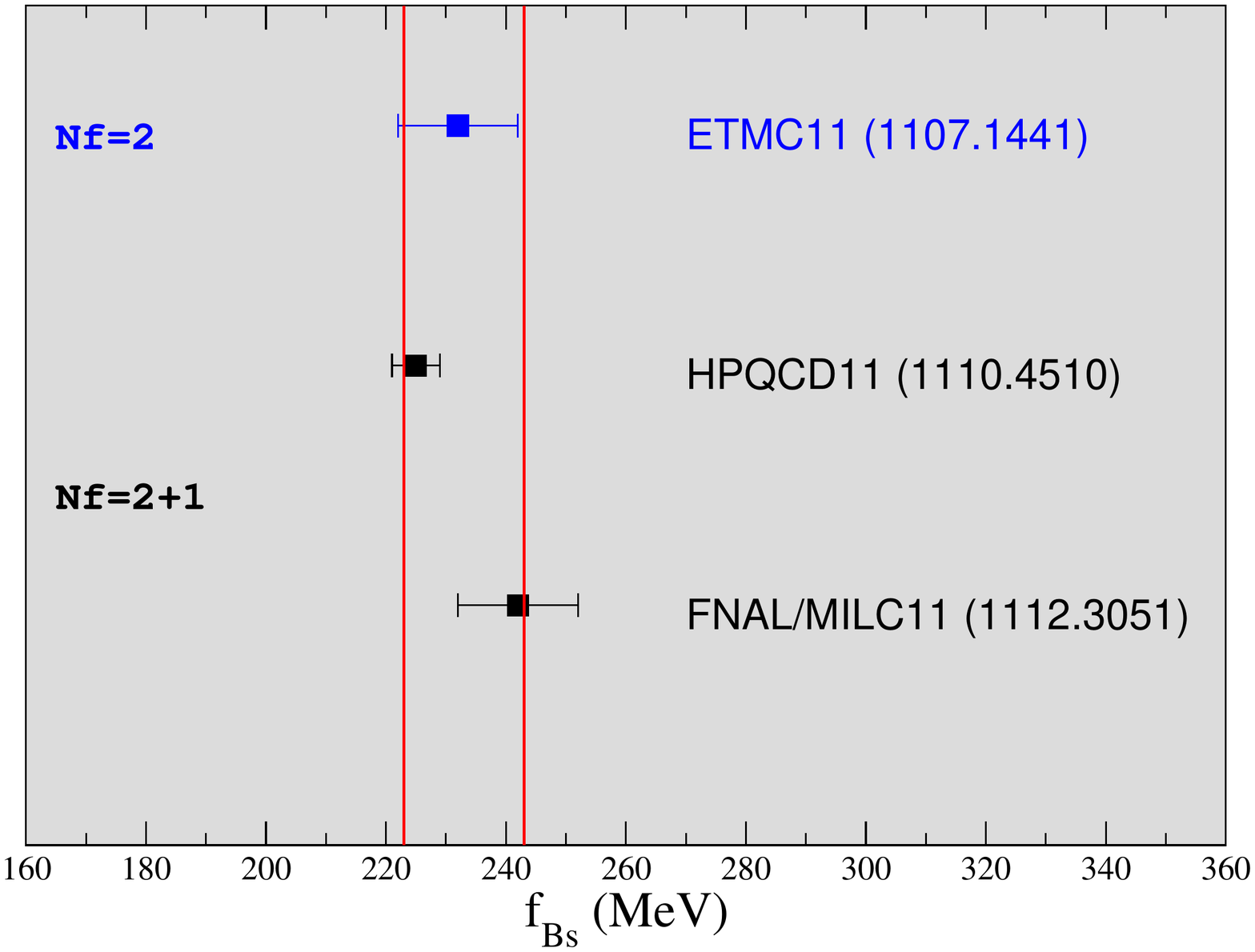} & 
\hspace*{-1.4cm}
\includegraphics[scale=0.31]{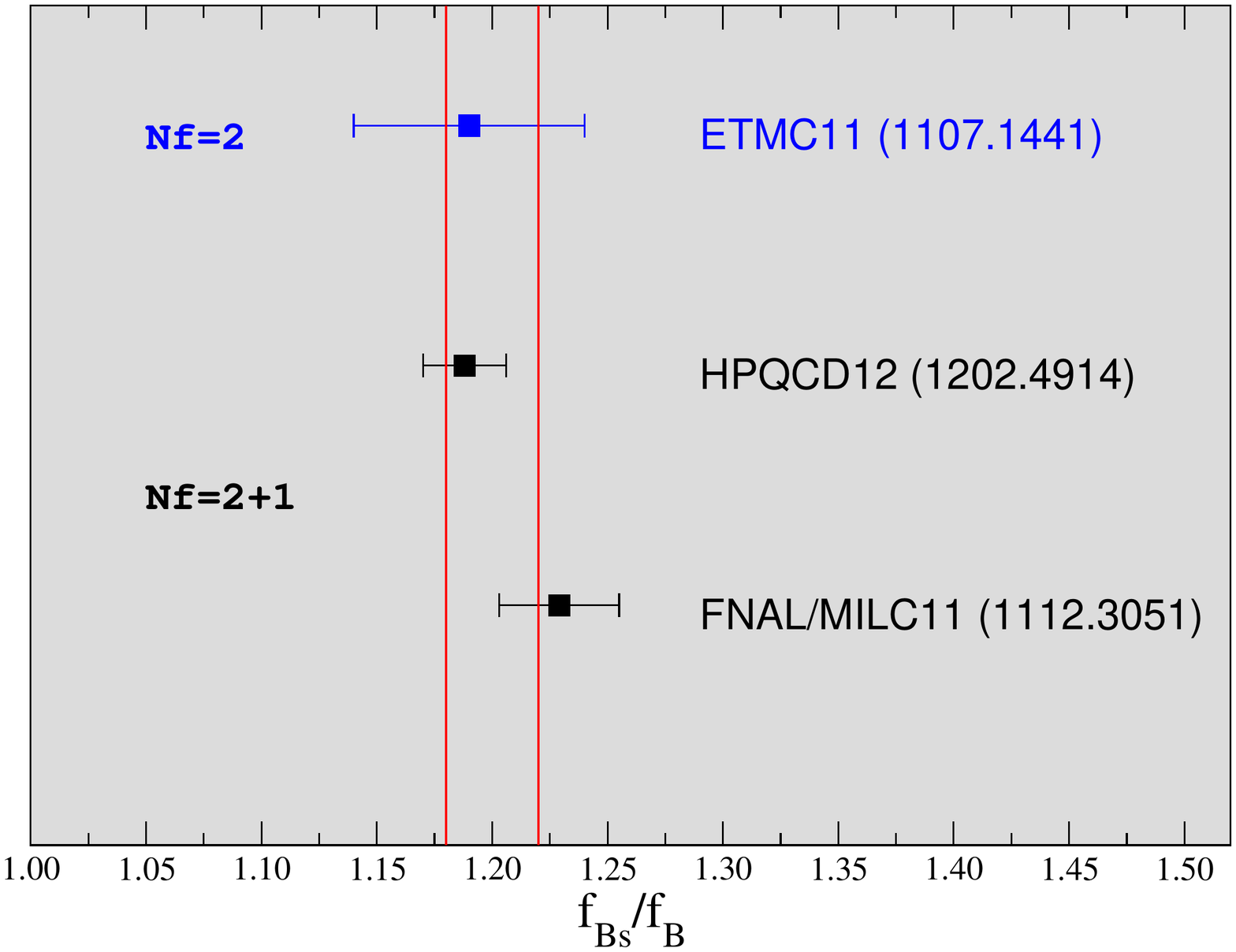}
\end{tabular}
\end{center}
\vspace{-1.0cm}
\caption{\sl Collection of recent unquenched results for $f_{Bs}$ (left) and $f_{Bs}/f_B$ (right). The red bands represent the average values used as input in the UTA.}
\label{fig:fBsfB}
\end{figure}

As far as the decay constants are concerned, it is convenient to consider $f_{Bs}$, which is almost insensitive to che chiral extrapolation as it depends on the light quark mass only in the sea, and the ratio  $f_{Bs}/f_B$ which has the advantage, w.r.t. $f_B$, of a partial cancellation of the statistical fluctuations and of the discretization effects.
Recent lattice results for $f_{Bs}$ and $f_{Bs}/f_B$~\cite{Dimopoulos:2011gx}-\cite{Na:2012kp} are collected in fig.~\ref{fig:fBsfB}. The average values adopted by the UTfit collaboration have been obtained as simple averages of these unquenched ($N_f=2$ and $N_f=2+1$) results, with a conservative error that corresponds to the ``typical'' accuracy of recent calculations. In particular, this means that the uncertainty on the $f_{Bs}$ average is larger than the error quoted for the HPQCD result~\cite{McNeile:2011ng}. The UTA inputs read
\beq
f_{Bs}=(233\pm 10)\mev\,\qquad f_{Bs}/f_B=1.20 \pm 0.02\,,
\eeqn
from which it also follows the average for $f_B$ given by
\beq
f_B=(194\pm 9)\mev\,.
\eeqn
New accurate analyses are being performed by FNAL/MILC, RBC/UKQCD, ETMC and Alpha and have been announced at the Lattice2012 conference~\cite{neil}-\cite{bernardoni}.

An accurate determination of the $B$ and $B_s$ decay constants is of great interest also for NP analyses. For instance $f_{Bs}$ enters quadratically in the theoretical prediction of the branching ratio for the $B_s \to \mu^+ \mu^-$ rare decay, which is highly sensitive to NP as it is a loop flavor changing neutral current process and it is theoretically clean being a leptonic decay.
The experimental bound has been recently strongly improved thanks to the LHC measurements. The combination of the the Atlas, CMS and LHCb bounds, that is dominated by the LHCb result, provides $Br(B_s \to \mu^+ \mu^-) < 
4.2 \cdot 10^{-9}$ at $95$\% CL~\cite{eerola}, which is close to the SM prediction. The SM prediction obtained from the UTA reads $Br(B_s \to \mu^+ \mu^-)_{SM} = (3.5 \pm 0.3) \cdot 10^{-9}$ and is in agreement with the recent result~\cite{Buras:2012ru}, $Br(B_s \to \mu^+ \mu^-)_{SM} = (3.23 \pm 0.27) \cdot 10^{-9}$, that has been obtained in a scheme such that NLO electroweak corrections are likely to be negligible and by paying attention to the effect of the soft radiation.

It is interesting to observe that Lattice QCD can provide useful information also for the experimental measurement of some processes like $B_s \to \mu^+ \mu^-$. Experimentally, in fact, it is fundamental to know the fragmentation fraction $f_s/f_d$ that is the fraction of b-quarks produced in the hadron collider that hadronize into $B_s$ mesons. Through factorization, the fragmentation fraction can be related to the ratio of the form factors involved in the semileptonic decays $B^0 \to D^+ \ell^- \bar \nu$ and  $B_s^0 \to D_s^+ \ell^- \bar \nu$. The FNAL/MILC collaboration has computed this ratio on the Lattice (with $N_f=2+1$ dynamical fermions and at two lattice spacings) finding $(f_s/f_d)=0.28 \pm 0.04$~\cite{Bailey:2012rr}, which represents  an improvement w.r.t. to a twenty-years-old QCD sum rule estimate~\cite{Blasi:1993fi}. 

The theoretical predictions for $B^0_{(s)}-\bar B^0_{(s)}$ require, in addition to the decay constants, the bag-parameters $B_B$ and $B_{Bs}$. It is convenient, as for the decay constants, to take $B_{Bs}$ and the ratio  
$B_{Bs}/B_B$ in input.
For these quantities the UTA lattice inputs coincide with the $N_f=2+1$ HPQCD results~\cite{Gamiz:2009ku}
\beq
\hat B_{Bs}=1.33\pm0.06\,\qquad B_{Bs}/B_B=1.05 \pm 0.07\,,
\eeqn
as other unquenched results are still preliminary~\cite{freeland, nuria}.
First unquenched results for the bag-parameters of the complete operator basis that describes $B^0_{(s)}-\bar B^0_{(s)}$ in NP models are also looked forward. An analysis by FNAL/MILC is in progress~\cite{freeland}.

For $|V_{ub}|$ and $|V_{cb}|$ there exist two different determinations based on the analysis of inclusive or exclusive semileptonic $B$ decays. The inclusive determination is in principle less affected by the non-perturbative uncertainties related to the hadronic final states. However, as the experimental inclusive measurements require the introduction of energy cuts, the inclusive determinations of $|V_{ub}|$ and $|V_{cb}|$  cannot avoid some model dependence in treating long-distance contributions at threshold.
This is not the case for the exclusive determinations which, instead, rely on theoretically clean lattice determinations of the form factors.

\begin{figure}[t!]
\begin{center}
\hspace*{-0.5cm}
\includegraphics[scale=0.31]{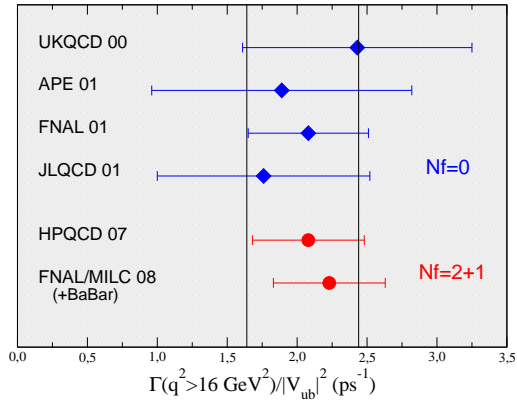}
\end{center}
\vspace{-1.0cm}
\caption{\sl Collection of quenched and unquenched results for $\Gamma(q^2>16\gev^2)/|V_{ub}|^2$. The red band represents the average considered by the UTfit collaboration.}
\label{fig:vub}
\end{figure}
For the exclusive determination of $|V_{ub}|$ on needs on the experimental side the measurement of the decay width for $B\to \pi \ell \nu$ and from Lattice QCD the hadronic quantity $\Gamma(q^2>16\gev^2)/|V_{ub}|^2$ (the large-$q^2$ region is more directly accessible to lattice determinations).
A collection of the available results for $\Gamma(q^2>16\gev^2)/|V_{ub}|^2$~\cite{Bowler:1999xn}-\cite{Bailey:2008wp} is provided in fig.~\ref{fig:vub}, where the red band corresponds to the average
\beq
|V_{ub}|_{excl}=(3.28 \pm 0.31)\cdot 10^{-3}\,,
\eeqn
which takes into account the (only) two modern unquenched results and older quenched results.
The comparison to the average quoted by the Heavy Flavor Averaging Group (HFAG)~\cite{hfag} for the inclusive determination, $|V_{ub}|_{incl}=(4.41\pm0.28)\cdot 10^{-3}$,  shows a $2.6 \sigma$ discrepancy, indicating that the $|V_{ub}|$ puzzle is still to be solved.
Further lattice calculations are certainly desired and are being performed by RBC/UKQCD~\cite{kawanai}, Alpha~\cite{bernardoni} and HPQCD~\cite{bouchard} or under investigation~\cite{gottlieb}.
The UTfit collaboration conservatively combines the two (exclusive and inclusive) values, using as UTA input $|V_{ub}|_{input}=(3.82\pm0.56)\cdot 10^{-3}$.
As we can see from table~\ref{tab:pull} the UTA prefers a value for $|V_{ub}|$ that is closer to the (lower) exclusive determination.

The state of the art for $V_{cb}$ presents a similarity to the $V_{ub}$ case. The inclusive determination derives from a global fit based on an Operator Product Expansion (OPE), in which $V_{cb}$ is fitted together with the b quark mass. The HFAG average~\cite{hfag} reads $|V_{cb}|_{incl}=(41.9\pm0.8)\cdot 10^{-3}$ and it is $2.4 \sigma$ larger than the exclusive value
\beq
|V_{cb}|_{excl}=(39.0 \pm 0.9)\cdot 10^{-3}\,.
\eeqn
We observe that the present accuracy on $|V_{cb}|$ is at the $2$\% level, that is approximately five times better than on $|V_{ub}|$. Both the inclusive and the exclusive determinations, in fact, are better under control for $|V_{cb}|$ than for $|V_{ub}|$. The reasons are the experimental cuts at higher energies, where the OPE is more reliable, for the inclusive determination, and the fact that the form factors involved in the exclusive determination of $|V_{cb}|$ measure a small deviation from the unity value in the infinite quark mass limit.

\begin{figure}[t!]
\begin{center}
\vspace*{-1.6cm}
\begin{tabular}{cc}
\hspace*{-1.0cm}
\includegraphics[scale=0.32]{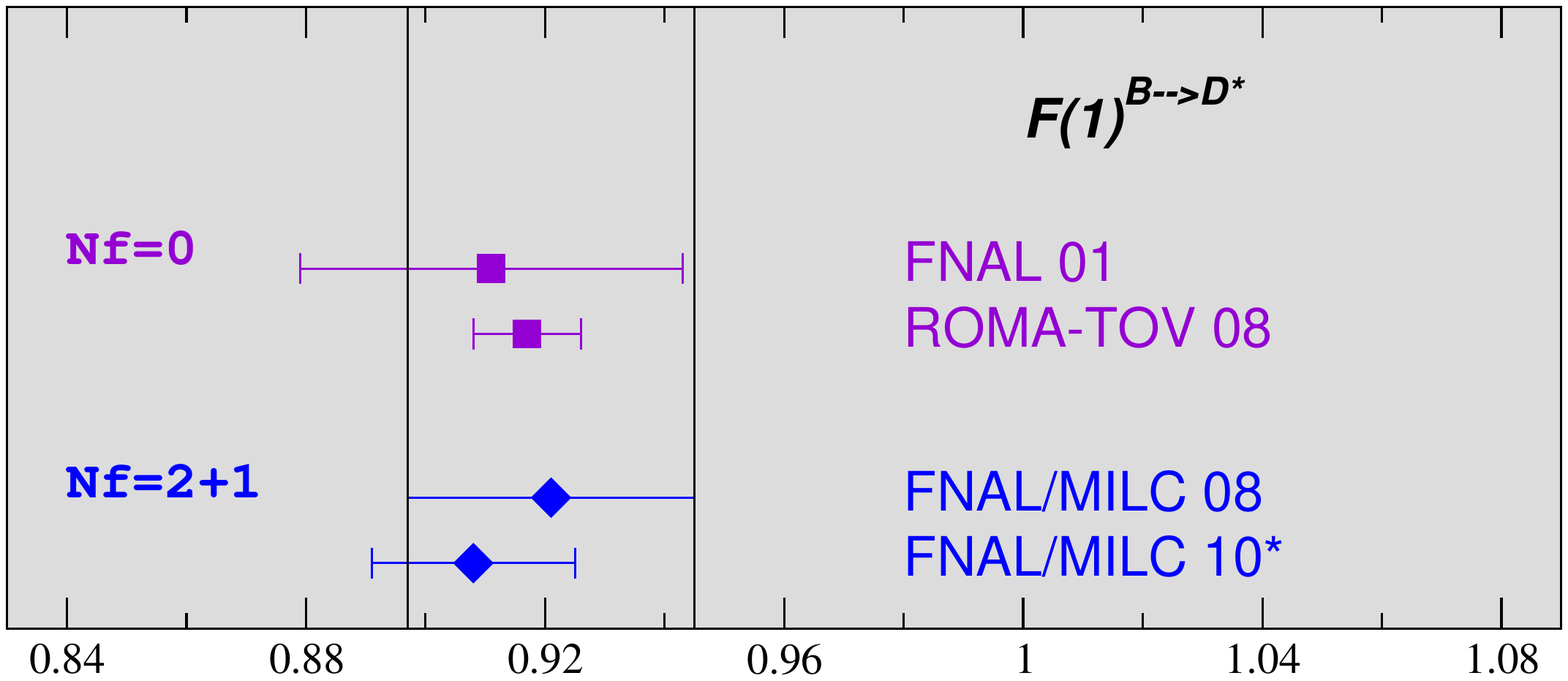} & 
\hspace*{-1.5cm}
\includegraphics[scale=0.32]{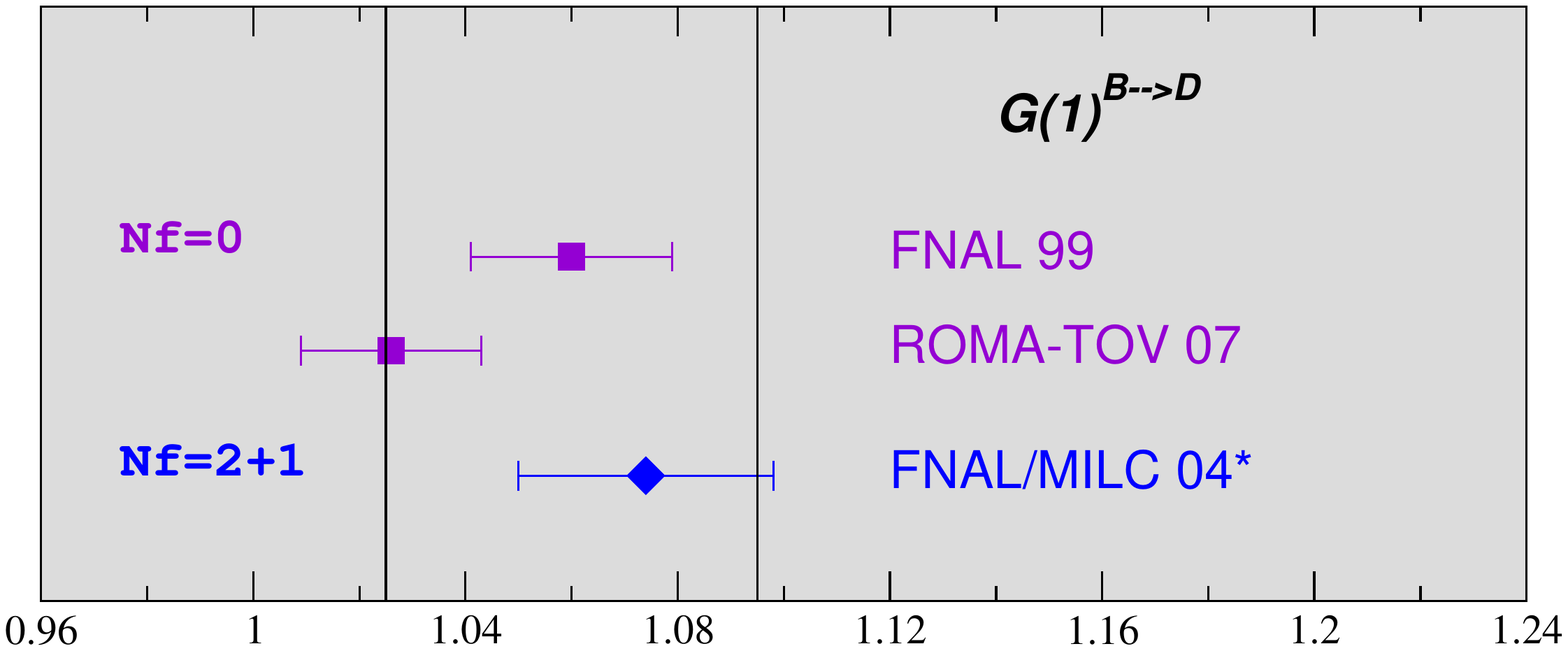}
\end{tabular}
\end{center}
\vspace{-1.0cm}
\caption{\sl Collection of quenched and unquenched results for the form factors $F(1)$ (left) and $G(1)$ (right). A star in the legends labels preliminary results.}
\label{fig:fg}
\end{figure}

Two channels are considered for the exclusive determination, $B \to D^* \ell \nu$ and $B \to D \ell \nu$,  which respectively require the lattice calculation of the form factors denoted as $F(1)$ and $G(1)$. Quenched and unquenched lattice results~\cite{Hashimoto:2001nb}-\cite{Bailey:2010gb} for $F(1)$ and $G(1)$ are shown in fig.~\ref{fig:fg}.
At present the $B \to D^*$ channel is measured with a better accuracy than the $B \to D$ channel, so that the exclusive determination of $V_{cb}$ relies on the lattice results for the form factor $F(1)$. Only one unquenched result~\cite{Bernard:2008dn} exists so far, more recently confirmed by FNAL/MILC itself at Lattice2010~\cite{Bailey:2010gb}.
The UTfit collaboration conservatively combines the exclusive and inclusive values, using as UTA input $|V_{cb}|_{input}=(41.0\pm 1.0)\cdot 10^{-3}$.
As we can see from table~\ref{tab:pull}, at variance with the $V_{ub}$ case, the UTA prefers a value for $|V_{cb}|$ that is closer to the (higher) inclusive determination.

\section{The UTA beyond the Standard Model}
The UTA, besides providing a strong tool for an accurate determination of the CKM parameters, can put constraints on possible NP effects.
To this purpose the UTfit collaboration performs the UTA without assuming the validity of the SM and parameterizing in a model independent way the NP effects that more probably might be visible, i.e. the NP contributions in meson-antimeson mixing phenomena ($K^0-\bar K^0$, $B^0-\bar B^0$, $B^0_s-\bar B^0_s$)~\cite{Bona:2005eu}. 
Including the new D0 data~\cite{Abazov:2009wg} and the recent LHCb measurement~\cite{LHCb:2012ad} for $B^0_s-\bar B^0_s$ mixing, the UTA beyond the SM finds that the NP effects in all three systems ($K^0-\bar K^0$, $B^0-\bar B^0$ and $B^0_s-\bar B^0_s$) are constrained to be compatible with zero~\cite{utfit}.
Further measurements are looked forward for the dimuon charge asymmetry, in order to confirm or discard the large (non-SM) value of the $B^0_s-\bar B^0_s$ mixing phase indicated by the D0 measurement~\cite{Hoeneisen:2011zz}.

The NP constraints provided by the UTA analysis beyond the SM can be converted into lower bounds on the NP scale.
Let us consider, for instance, the $K^0-\bar K^0$ system which at present provides the most stringent constraints on NP.
In models of physics beyond the Standard Model, the effective Hamiltonian that describes the $K^0-\bar K^0$ mixing amplitude involves in general the complete basis of $\Delta S=2$ four-fermion operators and it has schematically the form
\beq
\mathcal{H}_{eff}^{\Delta S=2} = \sum_{i=1}^{5} C_i O_i+\sum_{i=1}^{3} \tilde{C}_i \tilde{O}_i\,.
\label{eq:ds2}
\eeqn
A common choice for the basis is constituted by the operators
\bea
\label{eq:fullop}
&& O_1 = \bar s^{\alpha}\gamma_\mu(1 -\gamma_5)d^{\alpha} 
      \bar s^{\beta}\gamma_\mu(1 - \gamma_5)d^{\beta}\ , \nn \\
&& O_2 = \bar s^{\alpha}(1 - \gamma_5)d^{\alpha} 
      \bar s^{\beta}(1 - \gamma_5)d^{\beta}\ , \nn \\
&& O_3 = \bar s^{\alpha}(1 - \gamma_5)d^{\beta} 
       \bar s^{\beta}(1 - \gamma_5)d^{\alpha} \ , \\
&& O_4 = \bar s^{\alpha}(1 - \gamma_5)d^{\alpha} 
      \bar s^{\beta}(1 + \gamma_5)d^{\beta}\ , \nn \\
&& O_5 = \bar s^{\alpha}(1 - \gamma_5)d^{\beta} 
      \bar s^{\beta}(1 + \gamma_5)d^{\alpha}\ , \nn
\eea
where $\alpha$ and $\beta$ are color indices, together with the operators
$\tilde O_{1,2,3}$ obtained from $O_{1,2,3}$ with the exchange $\gamma_5 \to -
\gamma_5$. In chirally invariant renormalization schemes, the operators $\tilde
O_{i}$ have the same matrix elements of the $O_{i}$ and, for this reason, they
will not be mentioned in what follows.
We observe that in the SM only the operator $O_1$ appears in the $K^0-\bar K^0$ amplitude.

The Wilson coefficients appearing in $\mathcal{H}_{eff}^{\Delta S=2}$, in an effective theory approach, can be parameterized in the form
\begin{equation}
  C_i (\Lambda) = \frac{F_i L_i}{\Lambda^2}\, ,\qquad i=2,\ldots,5\, ,
  \label{eq:cgenstruct}
\end{equation}
where $F_i$ is the (generally complex) relevant NP flavor coupling,
$L_i$ is a (loop) factor which depends on the interactions that
generate $C_i(\Lambda)$, and $\Lambda$ is the scale of NP, i.e.\ the
typical mass of new particles mediating $\Delta S=2$ transitions. For
a generic strongly interacting theory with an unconstrained flavor
structure, one expects $F_i \sim L_i \sim 1$, so that the
phenomenologically allowed range for each of the Wilson coefficients
can be immediately translated into a lower bound on
$\Lambda$. Specific assumptions on the NP flavor structure
correspond to special choices of the $F_i$ functions. For example
Minimal Flavor Violation
(MFV) models~\cite{Buras:2000dm,D'Ambrosio:2002ex} 
correspond to $F_1=F_\mathrm{SM}$ and $F_{i \neq 1}=0$.

\begin{figure}[t!]
\begin{center}
\vspace{-1.0cm}
\includegraphics[scale=0.7]{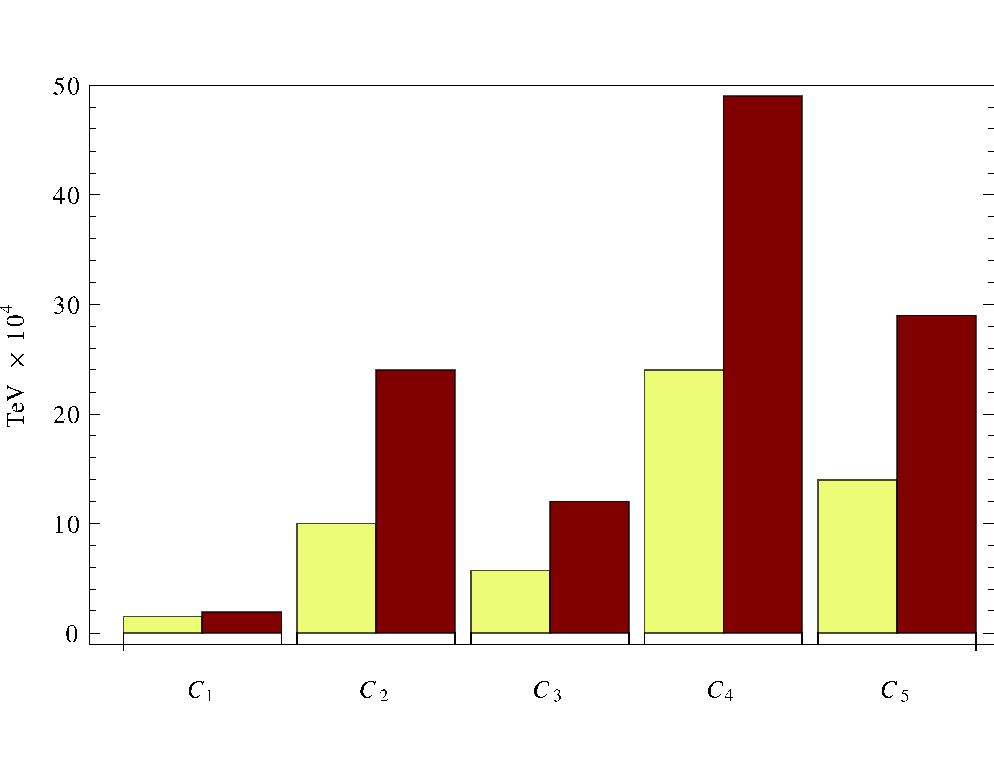}
\end{center}
\vspace{-1.0cm}
\caption{\sl Comparison of lower bounds on the NP scale between ref.~\cite{Bertone:2012cu} (red) and the older analysis~\cite{Bona:2007vi} (yellow).}
\label{fig:bounds}
\end{figure}

Updated lower bounds on $\Lambda$ have been recently obtained in ref.~\cite{Bertone:2012cu}, by following the same procedure of ref.~\cite{Bona:2007vi} and using the new unquenched lattice results for B-parameters of the complete basis operators calculated by ETMC~\cite{Bertone:2012cu} (with $N_f=2$ and three lattice spacings), which read~\footnote{For the definition of the matrix elements in terms of the B-parameters we refer to ref.~\cite{Bertone:2012cu}.}
\beq
B_2=0.54 \pm 0.03\,,\quad B_3=0.94 \pm 0.08\,,\quad B_4=0.82 \pm 0.05\,,\quad B_5=0.63 \pm 0.07\,,
\label{eq:bpar}
\eeqn
in the $\bar{MS}$ scheme defined in ref.~\cite{Buras:2000if} at a renormalization scale of $2 \gev$.
The ETMC results together with the new RBC/UKQCD results~\cite{Boyle:2012qb} (obtained with $N_f=2+1$ and at one lattice spacing) represent the first unquenched determination for $B_2-B_5$ and turn out to be in agreement. A further computation is being performed by the SWME collaboration~\cite{swme}.
The lower bounds on the NP scale as obtained from the constraints on the five Wilson coefficients are shown in fig.~\ref{fig:bounds} in a scenario of generic flavor structure with tree/strong NP interaction, compared to the older bounds~\cite{Bona:2007vi}.
The analysis, which in the considered scenario requires the NP scale to be larger than $\sim 5\cdot 10^5 \tev$, reflects the high sensitivity of Flavor Physics to NP effects.
To obtain the lower bound on
$\Lambda$ entailed by loop-mediated contributions, one simply has to
multiply the quoted bound by
$\alpha_s(\Lambda)\sim 0.1$ or $\alpha_W \sim 0.03$.

\section{Charm Physics}
In this section I discuss the charm sector of Flavor Physics, which is not involved in the UTA and can provide complementary information on the CKM matrix and on NP effects.
The phenomenon of $D^0-\bar D^0$ mixing, for instance, is sensitive to a different sector of NP w.r.t. $K$ and $B_{(s)}$ systems since $D$ mesons contains a charm quark, which is of up-type and might couple to different NP particles w.r.t strange and beauty quarks.
From the theory side $D^0-\bar D^0$ mixing has the disadvantage of being affected by large long-distance effects, due to the down and strange quarks circulating in the loop box diagrams, which dominate over the short-distance contribution.
Only order of magnitude estimates exist for the long-distance contributions and are at the level of the experimental constraints. This prevents the possibility of revealing an unambiguous sign of NP.
However, barring accidental cancellations between SM and NP contributions, significant constraints can still be put on the NP parameter space.

A very recent analysis of $D^0-\bar D^0$ mixing, which combines up to date experimental measurements, has been performed by the UTfit collaboration~\cite{:2012zkb}.
First (preliminary) unquenched lattice results for the five $B_D$-parameters entering $D^0-\bar D^0$ mixing in NP models, have been computed with $N_f=2$ dynamical flavors by ETMC~\cite{nuria}.
They read
\bea
&B_{D1}=0.77 \pm 0.05\,,\quad B_{D2}=0.73 \pm 0.04\,,\quad B_{D3}=1.37 \pm 0.11\,,&\nn\\
&B_{D4}=0.96 \pm 0.06\,,\quad B_{D5}=1.22 \pm 0.13\,,&
\label{eq:bDpar}
\eea
in $\bar{MS}$ at $2 \gev$.
These preliminary values have been used to constrain the parameter space of the Minimal Supersymmetric Standard Model (MSSM) with a generic Flavor structure~\cite{nuria}. The analysis is performed through the mass insertion approximation, that is in the so-called SuperCKM basis where squark-gluino couplings are flavor diagonal and the squark mass matrix $M_{\tilde u}^2$ is not.
The constraints on the $M_{\tilde u}^2$ off-diagonal elements turn out to be stronger than in the older analysis~\cite{Ciuchini:2007cw} by about a factor five, mainly thanks to the improved accuracy of the lattice results for the $B_D$-parameters and the decay constant $f_D$.

\begin{figure}[t!]
\begin{center}
\vspace{-2.0cm}
\begin{tabular}{cc}
\hspace*{-0.5cm}
\includegraphics[scale=0.31]{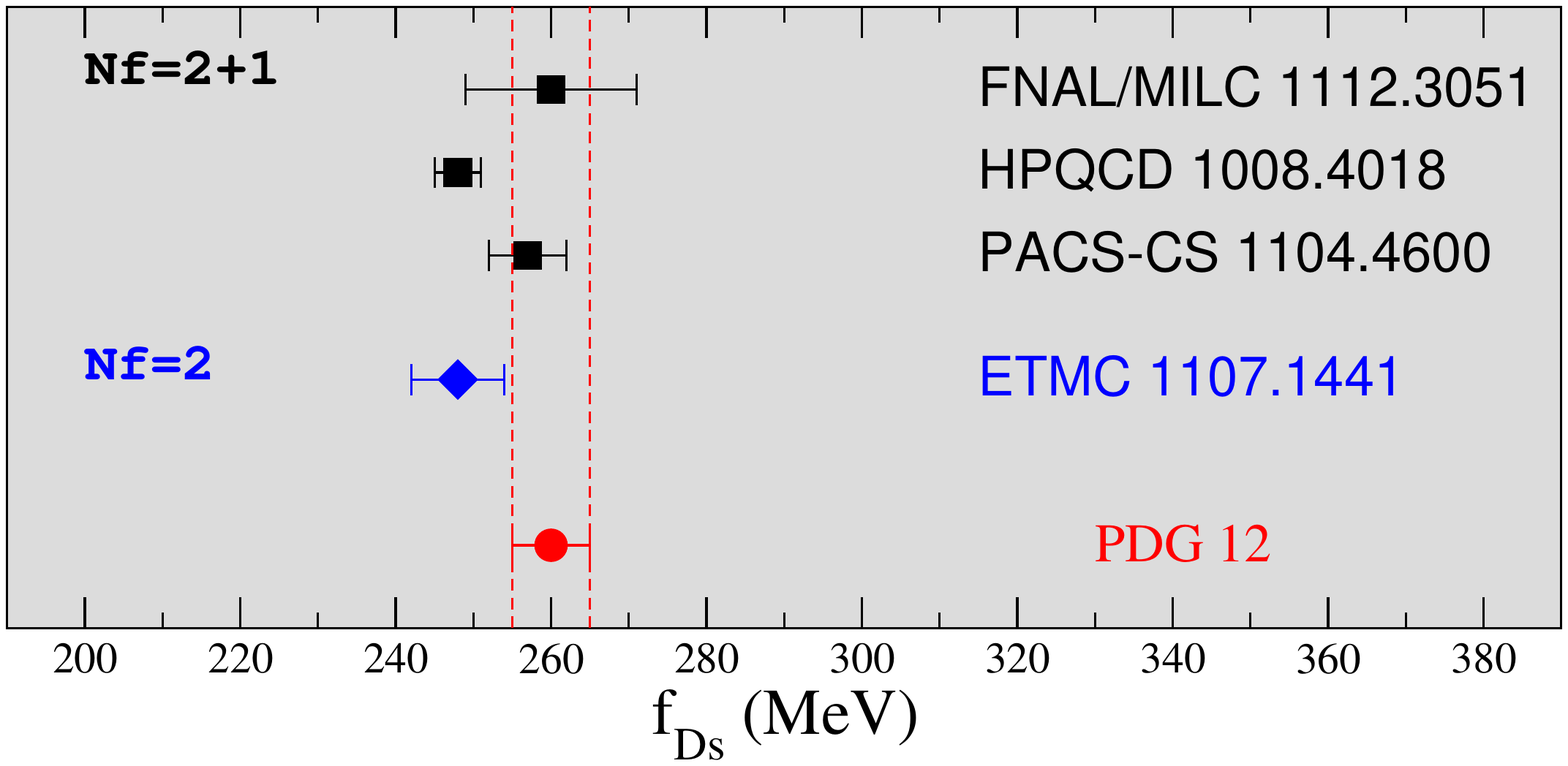} & 
\hspace*{-1.4cm}
\includegraphics[scale=0.31]{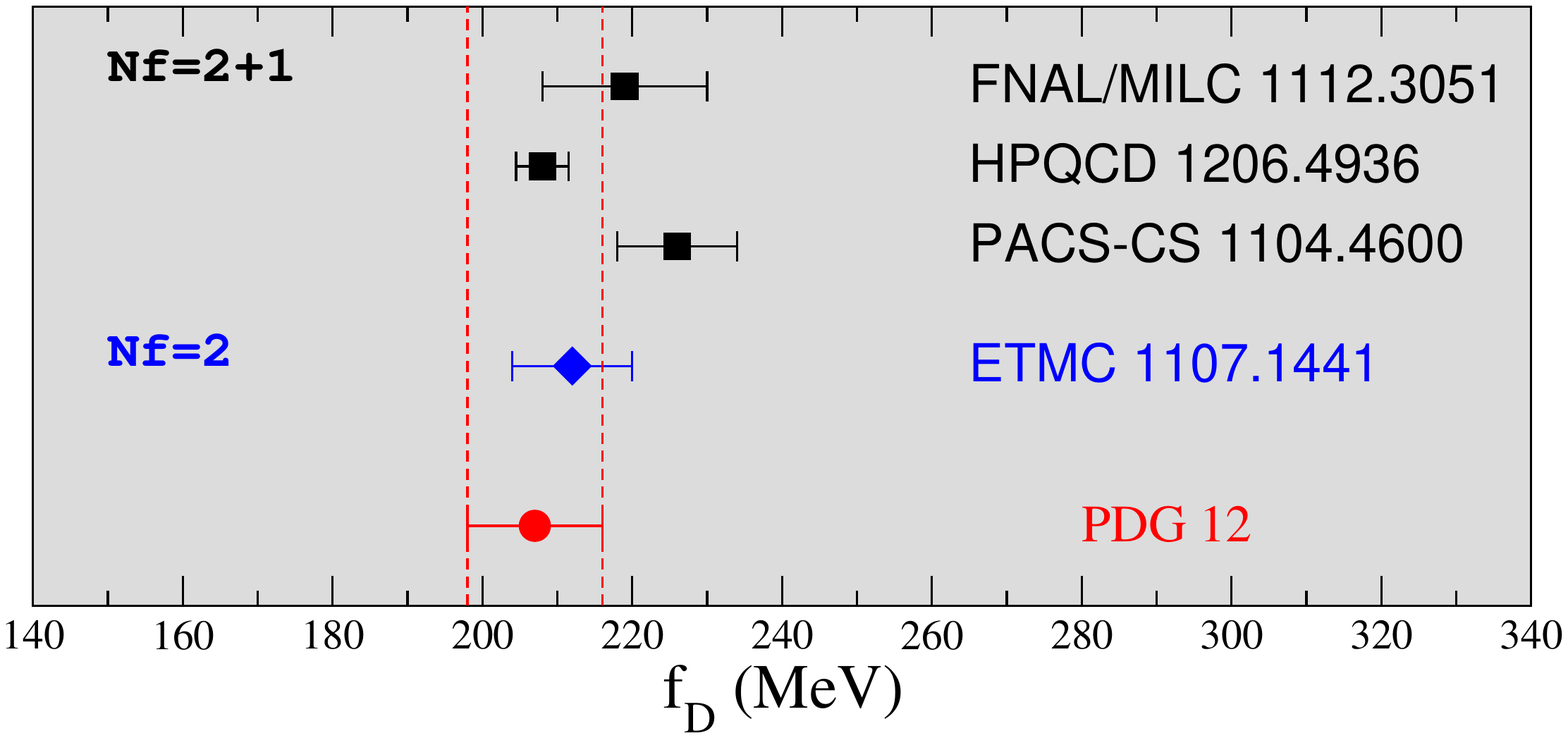}
\end{tabular}
\end{center}
\vspace{-1.0cm}
\caption{\sl Collection of recent unquenched results for $f_{Ds}$ (left) and $f_D$ (right). The red bands represent the PDG2012 averages.}
\label{fig:fdfds}
\end{figure}

For the pseudoscalar decay constants $f_D$ and $f_{Ds}$ several unquenched lattice calculations  have been performed in the last two years. A collection of these accurate results is shown in fig.~\ref{fig:fdfds}, with the red band representing the PDG value~\cite{Beringer:1900zz} obtained by assuming the CKM unitarity. Different lattice determinations turn out to be in good agreement and compatible to the unitarity prediction.

We recall that for $f_{Ds}$ in 2008 there was a deviation at the $3\sigma$ level between the HPQCD lattice result~\cite{Follana:2007uv} and the PDG experimental average~\cite{Amsler:2008zzb}. The HPQCD2007 result has been superseded by the HPQCD2010 analysis, where the determination of the scale has been improved~\cite{Davies:2010ip}, providing $f_{Ds}=(248 \pm 3)\mev$ that is $2.3\sigma$ higher than the older result. From the experimental side, new CLEO-c measurements have lowered the experimental average by $1.5\sigma$~\cite{Beringer:1900zz}.
As a consequence of these two effects the $f_{Ds}$ puzzle has gone away.
For $f_D$, HPQCD has studied the effect of improving the scale determination this year~\cite{Na:2012iu}, finding a results that is in very good agreement with the older one~\cite{Follana:2007uv} and slightly more accurate.
First calculations for the $D$ and $D_s$ decay constants that include the contribution of the charm quark in the sea (i.e. with $N_f=2+1+1$) are being performed by FNAL/MILC and ETMC. 

The lattice results for the decay constants $f_D$ and $f_{Ds}$ represent a fundamental ingredient in the determination of the CKM elements $V_{cd}$ and $V_{cs}$, respectively, from the study of $D$ and $D_s$ leptonic decays.
At present lattice and experimental uncertainties contribute at the same level.
An alternative determination of $V_{cd}$ and $V_{cs}$ comes from the $D\to \pi \ell \nu$ and $D\to K \ell \nu$ semileptonic decays, which require the lattice computation of the semileptonic form factors.
In this case the present lattice uncertainty dominates over the experimental one. The most accurate unquenched results for the $D \to \pi$ and $D \to K$ from factors have been obtained by HPQCD~\cite{Na:2011mc,Na:2010uf}, while a new improved analysis is being performed by FNAL/MILC~\cite{bailey}.

\section{A personal outlook to the future}
\begin{table}[t!]
\begin{center}
\begin{tabular}{||c|c|c||}
\hline
 & ICHEP2002~\cite{Lellouch:2002nj} & UTfit2012~\cite{utfit} \\
\hline
$\hat B_K$ & $0.86\pm0.15\quad [17\%]$ & $0.75\pm0.02\quad [3\%]$ \\
$f_{Bs} [\mev]$ & $238\pm 31\quad [13\%]$ & $233\pm 10\quad [4\%]$ \\
$f_{Bs}/f_B$ & $1.24\pm0.07\quad [6\%]$ & $1.20\pm0.02\quad [1.5\%]$ \\
$\hat B_{Bs}$ & $1.34\pm0.12\quad [9\%]$ & $1.33\pm0.06\quad [5\%]$\\
$B_{Bs}/B_B$ & $1.00\pm0.03\quad [3\%]$ & $1.05\pm0.07\quad [7\%]$ \\
$F(1)$ & $0.91\pm0.03\quad [3\%]$ & $0.92\pm0.02\quad [2\%]$ \\
$F_+^{B\to\pi}$ & $-\quad [20\%]$ & $-\quad [11\%]$ \\
\hline
\end{tabular}
\end{center}
\caption{Comparison between the lattice averages for the hadronic parameters entering the UTA, quoted by Laurent Lellouch at ICHEP2002 to the values used in input by UTfit in the 2012 analysis. Relative uncertainties are shown in square brackets.}
\label{tab:lattice0212}
\end{table}
In the present and next decades there will be a great experimental activity, not only in the direct NP searches at LHC, but also in the Flavor sector.
Within the quark sector the main role in Flavor Physics will be played by LHCb and the SuperB factories.
The latter experiments aim at improving the accuracy achieved at the B-factories by a factor $5-10$ and, in particular, at testing the CKM matrix at $1$\% level. 
They are also expected to increase the sensitivity for several channels of interest for NP searches by one order of magnitude.
Such experimental progress will require the control of the theoretical uncertainties, in particular of the lattice uncertainties on the hadronic parameters, at the same $1$\% level.
In order to try to understand if such a progress is feasible for Lattice QCD I briefly review the progress achieved in lattice calculations in the last ten years.
In table~\ref{tab:lattice0212} the lattice averages used in input at present in the UTA are compared to the lattice averages quoted by Laurent Lellouch in his review talk at ICHEP2002~\cite{Lellouch:2002nj}.
The comparison shows that an important progress has been achieved in Flavor Lattice QCD in the last ten years, which has typically led to a reduction of the uncertainties by a factor $2-5$.
This has mainly derived from the overcome of  the quenched approximation, made possible by the increase of the available computational power and better algorithms.
More recently further improvements are being realized, like simulations at the physical point, improved control of the discretization effects and the inclusion of the charm quark contribution in the sea.
I think that we can expect from Flavor Lattice QCD a further significant improvement in the next years, toward the $1$\% accuracy target.

\Acknowledgements
I wish to thank the organizers of the Charm2012, Lattice2012 and ICHEP2012 conferences for the invitation. I am grateful to Vittorio Lubicz for useful discussions and a careful reading of the paper.

\end{document}

%% file: econfmacros.tex
\def\beq{\begin{equation}}
\def\eeq#1{\label{#1}\end{equation}}
\def\eeqn{\end{equation}}

\def\beqa{\begin{eqnarray}}
\def\eeqa#1{\label{#1}\end{eqnarray}}
\def\eeqan{\end{eqnarray}}

\let\bar=\overbar

\def\Dslash{\not{\hbox{\kern-4pt $D$}}}
\def\dslash{\not{\hbox{\kern-2pt $\del$}}}

\def\msb{{\bar{\ssstyle M \kern -1pt S}}}

\def\s#1{\widetilde{#1}}
